\documentclass{article}
\def\1ad{\mbox{\normalsize $^1$}}
\def\2ad{\mbox{\normalsize $^2$}}
\def\3ad{\mbox{\normalsize $^3$}}
\def\4ad{\mbox{\normalsize $^4$}}
\def\5ad{\mbox{\normalsize $^5$}}
\def\6ad{\mbox{\normalsize $^6$}}
\def\7ad{\mbox{\normalsize $^7$}}
\def\8ad{\mbox{\normalsize $^8$}}
\def\makefront{
\vspace*{1cm}\begin{center}
\def\sp{
\renewcommand{\thefootnote}{\fnsymbol{footnote}}
\footnote[4]{corresponding author : \email_speaker}
\renewcommand{\thefootnote}{\arabic{footnote}}
}
\def\newtitleline{\\ \vskip 5pt}
{\Large\bf\titleline}\\
\vskip 1truecm
{\large\bf\authors}\\
\vskip 5truemm
\addresses
\end{center}
\vskip 1truecm
{\bf Abstract:}
\abstracttext
\vskip 1truecm
}
\usepackage{amssymb}
\def\beq{\begin{equation}}                     % 
\def\eeq{\end{equation}}                       %
\def\bea{\begin{eqnarray}}                     %         %
\def\eea{\end{eqnarray}}                       %       % 
\def\R{\mathbb R}
\def\C{\mathbb C}
\def\P{\mathbb P}
\def\S{\mathbb S}
\begin {document}                 
\begin{flushright}
\large \hfill{AEI-2003-005}\\
\hfill{hep-th/0301098}

\end{flushright}

\vspace{15pt}

\def\titleline{
%%%%%%%%%%%%%%%%%%%%%%%%%%%%%%%%%%%%%%%%%%%%%%
%                                            
% Insert now the text of your title.         
% Make a linebreak in the title with         
%                                            
%            \newtitleline                   
%                                            
%%%%%%%%%%%%%%%%%%%%%%%%%%%%%%%%%%%%%%%%%%%%%%%
Intersecting branes and 7-manifolds with $G_2$ holonomy\footnote{Contribution 
to the proceedings of the 35th Symposium Ahrenshoop, August 2002.}
%                                             %       %
%%%%%%%%%%%%%%%%%%%%%%%%%%%%%%%%%%%%%%%%%%%%%%%         %
}
\def\email_speaker{
{\tt 
%%%%%%%%%%%%%%%%%%%%%%%%%%%%%%%%%%%%%%%%%%%%%%
%                                                  
% Insert now the e-mail address of the speaker or  
% the author that should get the electronic mail   
% of the publishing house                           
%                                                  
%%%%%%%%%%%%%%%%%%%%%%%%%%%%%%%%%%%%%%%%%%%%%%         %
behrndt@aei.mpg.de
%                  %     %%%%%%%%%%%%%%
%                                            %       %       
%%%%%%%%%%%%%%%%%%%%%%%%%%%%%%%%%%%%%%%%%%%%%%         %
}}
\def\authors{
%                                            
%  Insert now the name (names) of the author 
%  (authors).                                
%  In the case of several authors with       
%  different addresses use labels e.g.       
%                                            
%             \1ad  , \2ad  etc.             
%                                            
%  to indicate that a given author has the   
%  address number 1 , 2 , etc.               
%  Signify the person with the above e-mail  
%  address with \sp (behind the address      
%  labels, e.g.                              
%                                            
%            \2ad\sp                         
%                                            
%  when the 2nd author                       
%  happens to be the 'speaker')                                  
%%%%%%%%%%%%%%%%%%%%%%%%%%%%%%%%%%%%%%%%%%%%%%         %
Klaus Behrndt 
%                                            %       %
%%%%%%%%%%%%%%%%%%%%%%%%%%%%%%%%%%%%%%%%%%%%%%         %
}
\def\addresses{
%%%%%%%%%%%%%%%%%%%%%%%%%%%%%%%%%%%%%%%%%%%%%%
%                                            
% List now the address. In the case of       
% several addresses list them by numbers     
% using e.g.                                 
%                                            
%             \1ad , \2ad   etc.             
%                                            
% to numerate address 1 , 2 , etc.           
%                                            
%%%%%%%%%%%%%%%%%%%%%%%%%%%%%%%%%%%%      %          
{\em Albert Einstein Institut \\            %          
Am M\"uhlenberg 1, 14476 Golm, Germany\\   %         %
behrndt{@}aei.mpg.de}
%%%%%%%%%%%%%%%%%%%%%%%%%%%%%%%%%%%%%%%%%%%%%
}
\def\abstracttext{
%%%%%%%%%%%%%%%%%%%%%%%%%%%%%%%%%%%%%%%%%%%%%
%                                             
% Insert now the text of your abstract.       
%                                             
%%%%%%%%%%%%%%%%%%%%%%%%%%%%%%%%%%%%%%%%%%%%%         %
In this talk I discuss intersecting brane configurations coming from
explicit metrics with $G_2$ holonomy.  An example of a 7-manifold
which representing a $\R^3$ bundle over a self-dual Einstein space is
described and the potential appearing after compactification over
the 6-d twistor space is derived.
%                       
%%%%%%%%%%%%%%%%%%%%%%%%%%%%%%%%%%%%%%%%%%%%%         %
}
\large
\makefront
%%%%%%%%%%%%%%%%%%%%%%%%%%%%%%%%%%%%%%%%%%%%%%%%
%                                              %
%  Insert now the remaining parts of           %
%  your article.                               %
%                                              %
%%%%%%%%%%%%%%%%%%%%%%%%%%%%%%%%%%%%%%%%%%%%%%%%

\section{Introduction}

The compactification of heterotic string theory on a complex
three-dimensional Calabi-Yau space was for a long time the standard
way to derive $N=1$ Super Yang Mills (SYM) in 4 dimensions (or even
better the MSSM) from string- or M-theory. In this approach
non-Abelian gauge groups as well as chiral matter appear very
natural. On the other hand in the dual type II or M-theory picture
non-Abelian gauge groups and chiral matter come from singularities or
appear on the world volume of D-branes. Concrete $N=1$ models could be
constructed using intersecting branes at angles, where supersymmetry
requires that the branes have to intersect at specific angles
\cite{150}. Examples are discussed in \cite{160} and
non-supersymmetric brane world models can be found in
\cite{360}, with massless matter living on the common
intersection of the branes.  One can use different branes to build the
brane world models, but especially interesting are the 6-branes which
uplift to pure geometry in the 11-d M-theory picture. Assuming that
the resulting 4-d external space is given by the Poincare invariant
flat Minkowski space, the 11-d geometry becomes: $M_{11} = M_4 \times
M_7$. Now, the amount of 4-d supersymmetry is directly related to the
number (covariantly constant) Killing spinors on $M_7$, which in turn
reduces the holonomy of this space. The holonomy of a generic
(orientable) 7-d space is given by $SO(7)$ and reduces to $G_2$ if the
space allows for exactly one covariantly constant spinor. If the space
allows for 2, 4 or even 8 Killing spinors the holonomy is further
reduced to $SU(3)$, $SU(2)$ or becomes trivial, i.e.\ consist only of
the identity. We will be interested in the case with only one Killing
spinor and thus having a $G_2$ manifold we want to address the
question of how one can obtain branes and the corresponding gauge
group upon dimensional reduction, see also \cite{240,230}.

\section{Branes from geometry}

Dp-branes are extended BPS objects that are charged under a (p+1)-form
potential.  {From} the Kaluza-Klein reduction one can obtain only a
1-form gauge potential supporting either a particle (0-brane) or the
Hodge-dual a co-dimension 3 brane. In 10 dimensions this is the
D6-brane which is obtained by the dimensional reduction from $M_7 =
R_3 \times X_{TN}$, where $X_{TN}$ is the 4-d Taub-NUT space with
the metric
\beq
ds^2 = {1 \over V} \big(d\chi + n \cos\theta \, d\varphi)^2 + V (dr^2 +
	r^2 d\Omega_2)\ ,
\quad  V = 1 + {n \over r} \ .
\eeq
{From} this metric the 6-brane with charge $n$ is obtained by
dimensional reduction along the Killing vector
$k=\partial_\chi$. Equivalently, a charge-$n$-6-brane can be seen as
$n$ 6-branes put on top of each other and hence there is a $U(n)$
world-volume gauge group.  Keeping the $4\pi$-periodicity of $\chi$ as
for the single brane, this results in a conical singularity, which is
nothing but the well known $\mathbb Z_n$ orbifold singularity. For
this simple case it is straightforward to extract the location and
charge of the brane. In fact, often one defines the space in terms of
orbifold actions, which gives a clear brane content (as orbifold fixed
points). Sometimes however, one wants to read-off the location and
charge of the 6-branes directly from a given 7-d metric where the
explicit orbifold action is unclear. Of course, branes appear only
after dimensional reduction along a given Killing vector $k$ and one
finds the identification
\begin{center}
\begin{tabular}{lll}
location of 6-branes & --- & fixed point set $L$ (i.e.\  $|k|^2 =0$)
of codimension four 
               \\
charge of 6-branes & --- & (inverse) surface gravity $\kappa$ of $L$ (
	$\kappa^2 = |\nabla k|^2$) \ .
\end{tabular}
\end{center}
For the case discussed above one finds $|k|^2 = {1 \over V}$ and
$|\nabla k|^2 = {1\over n} + {\cal O}(r)$ and the relations are
identical fulfilled. It is a well-known fact from black hole solutions
that a non-vanishing surface gravity (=non-vanishing Hawking
temperature) corresponds to case where the Killing vector is compact
(periodic Euclidean time) and the fixed point set is at finite
geodesic distance. In the extreme limit the surface gravity vanishes,
the Killing vector becomes non-compact or a translational isometry and
the fixed point moves to infinity.  For the simple 6-brane
configuration from above, there is no regular extreme limit, but note,
that if $\kappa =1$ there is no conical singularity and the geometry
is locally equivalent to the flat space. Since $n$ is an integer, one
can resolve the conical singularity by going to the $n$-center
solution and because $g_{\chi\chi}$ vanishes at each center, between
each two centers appear a non-trivial 2-cycle and the resolution of the
singularity can also be understood as the blowing up of these
2-cycles.  This is of course well-known for 6-branes living in flat
space, where the corresponding multicenter solution exists, but it may
be not possible for a more general manifold with $G_2$ holonomy (due
to the lack of the corresponding moduli).

Let us also mention, that in addition to co-dimension four
singularities, which are interpreted as 6-branes, there can also
appear singularities of co-dimension two and six. In general their
interpretation is less clear. For the example that we will discuss
below, co-dimension two singularities do not appear and the
co-dimension 6 objects are T-dual to NS5-branes.

We can also extract the 10-d quantities in detail. Having a Killing
vector $k=\partial_\chi$ we write the 11-d metric as usual
\beq
ds^2 = e^{{4\over 3} \phi} ( d\chi + C_\mu dx^\mu)^2 +
	e^{-{2\over 3}\phi} ds_{10}^2 
\eeq
and the 10-d fields can be expressed only by the Killing vector as
\beq e^{{4 \over 3} \phi} = |k|^2 \ , \quad C_{\mu} ={k_\mu \over
|k|^2} \ , \quad F_{\mu\nu} = {6 k^\alpha k_{[\alpha} \partial_\mu
k_{\nu]} \over |k|^2} \ .  
\eeq 
For the simple example above we find $\phi \sim \log |k|^2 =- \log V$
and the RR-1-form becomes $C = n \cos\theta \, d\varphi$ so that $n$
is in fact related to the 6-brane charge.

\section{Seven-manifolds with $G_2$ holonomy}

Recall, 7-manifolds with $G_2$ holonomy allow for exactly one
covariantly constant Killing spinor which can be used to build
covariantly constant differential forms as $F_{n_1 \cdots n_p} =
\bar\epsilon \gamma_{n_1\cdots n_p} \epsilon$.  If $\epsilon$ is a
commuting pseudo Majorana spinor (with a symmetric charge conjugation
matrix) on the 7-d space these $p$-forms are non-vanishing for
$p=0,3,4,7$ \cite{320}. The 0- and its dual 7-forms are trivial from
the geometry point of view, the only nontrivial case is given by the
covariantly constant 3-form (and its dual 4-form). In a proper
parameterization, this form is given by
\beq
\begin{array}{rcl}
\Phi = 
{1 \over 3!} \phi_{abc} \, e^a \wedge e^b \wedge e^c &=&
e^1 \wedge e^2 \wedge e^3 + e^4 \wedge e^3 \wedge e^5 +
e^5 \wedge e^1 \wedge e^6 + e^6 \wedge e^2 \wedge e^4 + \\[2mm] &&
e^4 \wedge e^7 \wedge e^1 +e^5 \wedge e^7 \wedge e^2 +
e^6 \wedge e^7 \wedge e^3 \ , 
\\[2mm]
&=& e^1 \wedge e^2 \wedge e^3 + e^i \wedge e^m J^i_{mn}\wedge e^n 
\end{array}
\eeq
where $e^a$ ($a=1...7$) are the vielbeine of the 7-manifold and
$J^i_{mn}$ is the triplet of anti-selfdual complex structures
satisfying the quaternionic algebra
\beq
J^i \cdot J^j = - {\mathbb I} \, \delta^{ij} + \epsilon^{ijk} J^k \ .
\eeq
The appearance of a triplet of complex structures suggests an
embedding of a hyper Kaehler or quaternionic space in the 7-manifold.
In fact, two major classes of $G_2$ manifolds are \cite{210}: (i)
an $\mathbb R^3$ bundle over a quaternionic space $Q$ or (ii) an
$\mathbb R^4$ bundle over $\mathbb S^3$ (where $\mathbb R^4$ can be
replaced by a more general hyper Kaehler space \cite{200}).  For both
cases exist a limit in which the space becomes a cone over a 6-d space
$Y_6$ so that the metric can be written as
\beq
ds^2_7 = d\rho^2 +\rho^2 ds_Y^2 \ .
\eeq
In case (i) $Y_6$ is the twistor space related to the quaternionic
base space (e.g.\ $Y_6={\mathbb{CP}^3}$ for $Q=\mathbb S^4$ or $Y_6 =
U(3)/U(1)^3$ for $Q=\mathbb{CP}_2$) and case (ii) gives $Y_6= \S^3
\times \S^3$. Obviously, $\rho = 0$ is a co-dimension 7 singularity,
where all branes will meet and this is exactly the point where chiral
matter is located \cite{230, 220}.  Viewed from the 10-d perspective
this matter corresponds to open strings stretched between different
branes, which become massless at the common intersection.  In 11
dimensions these open strings become membranes wrapping a 2-cycle
which terminates at the fixed point of the Killing vector. The kind of
matter depends crucially on the type of the singularity
\cite{230,220}, usually it transforms in the bi-fundamental
representation of the gauge group, but also matter transforming in a
tensor as well as tri-fundamental representation has been discussed
\cite{300}.  In any case, the number of chiral multiplets is related
to the second Betti number $b_2$ and for case (ii) with $Y_6=\S^3
\times \S^3$ we will get no chiral multiplets (there can be chiral
matter if $\R^4$ is replace by a more general hyper Kaehler space)
whereas for case (i) with $Q=\S^4$ there is one chiral multiplet and
for $Q=\C\P_2$ gives two chiral multiplets; more general quaternionic
spaces (see e.g.\ \cite{390,310,100,260}) will of course yield more
chiral multiplet.  The case with a single chiral multiplet (i.e.\ with
$Q=\S^4$) gives in 10 dimensions the situation with two 6-branes,
which are connected by the non-trivial 2-cycle. The case yielding two
chiral multiplets describes in 10 dimensions three 6-branes, which
exhibits a triality symmetry corresponding to the exchange of the
three 6-branes and thus all 6-branes have the same gauge group on
their world volume. More interesting is of course the situation with
different gauge groups where each singularity corresponds to a
different number of 6-branes. This brings us to the proposal to use a
$G_2$ manifold given as an $\R^3$ bundle over
$\mathbb{WCP}_{n_1n_2n_3}$, which has again $b_2=2$ and hence
describes three intersecting 6-branes with gauge groups related to the
weights $n_1$, $n_2$ and $n_3$ \cite{220}. This space however is known
to have further singularities, related to co-dimension 6 fixed points.

\section{The explicit example}

The mathematical literature \cite{260} provides strong evidence
that the metric of the metric of the quaternionic space
$\mathbb{WCP}_{n_1n_2n_2}$ is basically given by a fourth order
polynomial where the roots sum to zero, i.e.
\beq
F(x) = \kappa(x-r_1)(x-r_2)(x-r_3)(x-r_4) \quad , \qquad 
r_1 + r_2 + r_3 + r_4 = 0
\eeq
and can be written as
\beq \label{4dmetric}
ds_4^2 = {q^2 - p^2 \over F(p)} dp^2 + {p^2 - q^2 \over F(q)} dq^2
       + {F(p) \over q^2 -p^2} (d\tau + q^2 d\sigma)^2 
       + {F(q) \over p^2 -q^2} (d\tau + p^2 d\sigma)^2 
\eeq
and it solves the equations $R_{mn} = 3 \kappa g_{mn}$. Since the Weyl
tensor satisfies $W + {^{\star}W} =0$ these spaces are known in the
mathematical literature as (anti) self-dual Einstein spaces.  We
consider the compact case and write for the curvature parameter
$\kappa=1$, but one can of course consider any value. Due to a scaling
symmetry, this solution depends on three continuous and one discrete
parameter. In the physics literature this metric is known in Minkowski
signature ($q \rightarrow iq$ and $r_m \rightarrow i r_m$) as
E(A)dS-Kerr-Taub-NUT space \cite{250}, where the four parameters have
the interpretation as cosmological constant, the rotational parameter
and the mass equals the NUT parameter (due to the anti-self-duality
constraint). The remaining discrete parameter defines the slicing of
the 4-d space by a 2-space of constant positive, negative or vanishing
curvature.

The curvature square of this space is $R_{mnrs} R^{mnrs} = 24 \kappa^2
+ 96 n^2/(p+q)^6$, where $p+q=0$ is a non-removable singularity and $n
= \partial_x F|_{x=0}$ is the mass (=NUT parameter) for Minkowskean
solution. By a proper choice of parameter this singularity can however
be put into an unphysical coordinate region where the metric has the
signature (2,2). Note, the space has Euclidean signature only if the
two relations hold: $F(p) \geq q^2-p^2$ and $F(q) \geq p^2 -q^2$. With
this 4-d base space the 7-d metric reads
\beq \label{090}
ds^2 = {dr^2 \over (1 - {4u_0 \over r^4})} + {r^2 \over 4}
        \Big(1 - {4u_0 \over \, r^4}\Big) 
        \, h_{ab} (dx^a + \xi^a_i A^i\big) (dx^b + \xi^b_j A^j)
        + {r^2 \over 2} ds^2_4\ ,
\eeq
where $A^i = {1 \over 2} \omega^{mn} J_{mn}^i$ is the anti-self-dual
part of the spin connections $\omega^{mn}$ of the 4-d base space and
$\xi^a_i$ are three Killing vectors of the  $\S^2$-metric
$h_{ab}$; for more details see \cite{100,310}. In the limit where the
parameter $u_0$ vanishes, this metric describes in fact a cone over a
6-d space $Y_6$, which is topologically an $\S^2$ fibered over
the 4 dimensional base given by the metric (\ref{4dmetric}). 

This space has two commuting, tri-holomorphic\footnote{Which leaves
the triplet of $SU(2)$ connections invariant.} Killing vectors ($k_1 =
\partial_\sigma$ and $k_2 = \partial_\tau$) and taking a general
linear combination: $k= \beta_1 \partial_\tau - \beta_2
\partial_\sigma$, 6-branes are related to co-dimension four fixed
point sets. Recall, the fixed point set given by $|k|^2=0$ can have
different co-dimensions and one finds
\begin{center}
\begin{tabular}{rl}
co-dimension 7: &if $u_0=0$ at $r=0$ \ , \\
co-dimension 6: &at $|\xi|^2 =0$ and $F(p) = F(q) =0$ \ , \\
co-dimension 4: &if (i) $|\xi|^2 = 0$ and $q^2 = {\beta_1 \over \beta_2}
 = r_m$ (or $p^2 ={\beta_1 \over \beta_2}$) or \\
& if (ii) $F(q)=F(p)=0$ and $\beta_1 A_{\tau}^2 - \beta_2 A_\sigma^2 =0$ \ .
\end{tabular}
\end{center}
Explicit calculations show no co-dimension 2 fixed points (or better
they are at infinite geodesic distance).  The co-dimension 4
singularity has the interpretation as 6-branes and and the
co-dimension 7 case is related to the appearance of chiral
fermions. The co-dimension 6 singularity is related to a fixed point
of both Killing vectors and hence they correspond to additional
NS5-branes in the dual type IIB picture.  If one is interested in a
``clear'' picture of only 6-branes living in a topologically flat
space, one has to ``turn off'' the co-dimension 6 singularities and in
this case the number of components of the fixed point set gives the
number number of 6-branes \cite{290}.  This can be done by equalizing
two of the roots of the 4th order polynomial, e.g.\ $r_1 = r_2$ and
$k=r_3^2 \partial_\tau - \partial_\sigma$, see \cite{100}.

In concluding let us stress, because the 7-space is non-compact we
cannot simply reduce the model to 4 dimensions, but we can of course
reduce it over the compact 6-d space and obtain a domain wall solution in
5 dimensions. So, writing the 7-metric as
\beq 
ds^2 = e^{2\varphi_1(r)} dr^2 + e^{2\varphi_2(r)} 
        \, h_{ab} (dx^a + \xi^a_i A^i\big) (dx^b + \xi^b_j A^j)
        + e^{2\varphi_3(r)} ds^2_4\ .
\eeq
The Kaluza-Klein scalars $X_{2,3}$ parameterize the volumes of
the $\S^2$ and the quaternionic space and using well-known reduction
formulae \cite{330} it is straightforward to derive their 5-d potential.
It is obtained from the Ricci tensor by setting $r=constant$ and reads
\beq
V = 2\, e^{-{4\over 3} (\varphi_2 +2 \varphi_3)}\,
\Big( e^{-2\varphi_2} + 6 \, \kappa \, e^{-2\varphi_3} + 1\Big)
\eeq
where the first dilatonic term is due to the conformal rescaling
necessary to obtain the Einstein frame and we used that the 2- and
4-dimensional spaces satisfy $R_{ab}^{(2)} = h_{ab}$ and $R^{(4)} = 3
\kappa g_{mn}$, where we set $\kappa =1$ to have a compact space.
Like the potentials coming from flux compactifications \cite{400}
this potential has no fixed points and hence the scalars cannot be
stabilized yielding a singular domain wall.  An extrema for the
potential would of course imply that the solution would contain either
an AdS space or becomes flat, which is not the case for the explicitly
solution given in eqs.\ (\ref{090}). Similarly to the approach
discussed in \cite{500} this model 
should be embeddable into $N$=2,$D$=5 gauged supergravity.

\vspace*{10mm}

%%%%%%%%%%%%%%%%%%%%%%%%%%%%%%%%%%%%%%%%%%% 
{\bf Acknowledgment} I would like to thank Gianguido Dall'Agata,
Dieter L\"ust and Swapna Mahapatra for the collaboration on this
subject. This work is supported by a Heisenberg grant of the DFG.

%%%%%%%%%%%%%%%%%%%%%%

\newpage 

%%%%%%%%%%%%%%%%%%%%%%%%%%%%%%%%%%%%%%%%%%%%%%%%%%%%%%%%%%%%%%

% ---- Bibliography ----

% \nocite{*}                   %this uses *everything* in the .bib file
% \bibliography{proc}     %or whatever your .bib file is
% \bibliographystyle{utphys}   %if you use utphys.bst

\providecommand{\href}[2]{#2}\begingroup\raggedright\endgroup

\end{document}